\newcommand{\beq}{\begin{equation}}
\newcommand{\eeq}{\end{equation}}

\newcommand{\bs}{\thickspace}

\documentclass{emulateapj}
\usepackage{amssymb,amsmath,amstext,amsgen,amsopn,amsxtra,indentfirst}
\input psfig.sty

\slugcomment{ApJ, in press}

\shorttitle{Forming terrestrial planets in the presence of giant planet
scattering}
\shortauthors{Veras \& Armitage}

\begin{document}

\title{Predictions for the correlation between giant and terrestrial extrasolar 
planets in dynamically evolved systems}

\author{Dimitri Veras\altaffilmark{1,2} and Philip J. Armitage\altaffilmark{1,2}}
\altaffiltext{1}{JILA, Campus Box 440, University of Colorado, Boulder CO 80309-0440; 
dimitri.veras@colorado.edu; pja@jilau1.colorado.edu}
\altaffiltext{2}{Department of Astrophysical and Planetary Sciences, 
University of Colorado, Boulder CO 80309-0391}

\clearpage
\pagebreak
\newpage

\begin{abstract}
The large eccentricities of many giant extrasolar planets may represent the endpoint 
of gravitational scattering in initially more crowded systems. If so, the early 
evolution of the giant planets is likely to be more restrictive of terrestrial 
planet formation than would be inferred from the current, dynamically 
quiescent, configurations. Here, we study statistically the extent of the anti-correlation 
between giant planets and terrestrial planets expected in a scattering model. We 
use marginally stable systems of three giant planets, with a realistic range of 
planetary masses, as a simple model for the initial conditions prior to scattering, 
and show that after 
scattering the surviving planets reproduce well the known extrasolar planet eccentricities 
beyond $a > 0.5$~AU. By tracking the minimum periastron values of all planets during 
the evolution, we derive the distribution of orbital radii across which strong 
perturbations (from crossing orbits) are likely to affect low mass planet formation. 
We find that scattering affects inner planet formation at orbital separations less
than 50\% of the final periastron distance, $q_{fm}$, of the innermost massive 
planet in approximately 30\% of the realizations, and can occasionally 
influence planet formation at orbital separations less than 20\% of $q_{fm}$. 
The domain of influence of the scattering massive planets increases 
as the mass differential between the massive planets decreases. Observational 
study of the correlation between massive and terrestrial extrasolar planets 
in the same system has the potential to constrain the origin of planetary 
eccentricity.
\end{abstract}

\keywords{solar system: formation --- planets and satellites: formation --- 
planetary systems: formation --- celestial mechanics}

\section{Introduction}
Studies of the population of small bodies -- comets, asteroids, and Kuiper belt 
objects -- have provided powerful constraints on the early dynamical evolution 
of the massive planets in the Solar System. The concentration of objects in 
the outer Solar System in narrow zones in resonance with Neptune, with 
significant eccentricities, provides evidence for an early expansion of 
Neptune's orbit \citep{malhotra95,hahn99,murrayclay05}, and has motivated a scenario 
in which the orbits of Saturn, Uranus and Neptune all expanded as a 
consequence of scattering the planetesimal debris left over after 
planet formation \citep{thommes99,thommes02,tsiganis05}. Migration of 
Jupiter -- albeit for a much smaller distance -- is also predicted 
within such models and may have left its own signature in the 
distribution of Hilda asteroids \citep{franklin04}.

There is no near or medium term prospect for obtaining observations that 
would permit directly comparable studies in extrasolar planetary systems. 
A wealth of circumstantial evidence, however, suggests that the early evolution 
of many extrasolar planetary systems may have been more stochastic than 
in the Solar System, and these more dramatic histories {\em can} be 
investigated by studying the relationship between terrestrial planets (which 
are here being utilized purely as test particles of negligible mass)
and giant planets within the same system. The space-based transit mission 
{\em Kepler} \citep{basri05} will offer the first definite opportunity to 
conduct such investigations, though terrestrial planets may also be 
discovered via giant planet transit timing techniques \citep{agol05,holman05} 
or high-precision astrometry with the {\em Space Interferometry Mission} \citep{ford03b}. 
Gravitational microlensing also has potential for terrestrial planet discovery 
\citep{bennett02}, although the prospects for determining the overall architecture 
of systems in which planets are discovered are poor with this method.

High eccentricities have become a common signature of known extrasolar planets 
which reside outward of the tidal circularization limit. The origin of these 
eccentricities remains unclear. Further, \cite{goldreich04} 
have suggested that the giant planets in the Solar System might have 
experienced large eccentricities before becoming circularized to their
present values.  One popular model attributes the eccentricity 
to gravitational scattering among multiple massive planets \citep{rasio96,
lin97,levilissdunc98,ford03a,marzari02}. Scattering simulations have produced a 
quantitative match to the distribution of known extrasolar planet 
eccentricities \citep{ford03a}, and can also explain the current configuration 
of the upsilon Andromedae system \citep{ford05}. 
However, we cannot rule out the possibility that protoplanetary disk physics
allows planetary eccentricity to be excited 
via the interaction between a {\em single} massive planet and a gas 
disk \citep{artymowicz91,papaloizou01,ogilvie03,goldreich03}.

These two models for eccentricity growth have different consequences for the 
formation of terrestrial planets within the same system. If eccentricity arises 
from planet-disk interactions, then an eccentricity comparable to the final 
value is likely to be established soon after the formation of the giant planet, 
which must occur prior to the dissipation of the gas disk during the first 
few Myr \citep{haisch01}. This schedule probably precedes the final stages of terrestrial 
planet formation. \cite{chambers02}, \cite{leviagno03} and \cite{mardling04} have studied  
in detail the effect of the giant planet configuration, and its evolution, on the 
formation of terrestrial planets. If, conversely, eccentricity is the endpoint 
of gravitational scattering among several massive planets, then the 
dynamics involved in the scattering process will usually involve multiple 
rearrangements of the planets prior to stabilization (usually by ejection) and 
establishment of a `final' configuration. The extent of the ejection is dependent
on the masses, radii, and initial orbital separations of the planets.
During such rearrangements, which can 
take many Myr, giant 
planets can temporarily intrude (either physically, or via their resonances) 
into the terrestrial planet zone. The resulting constraints on where terrestrial 
planets can form and survive are bound to be more restrictive than those 
calculated assuming that the current observed (i.e. final) configuration of massive 
planets has existed for all time \citep{jonesleecham01,menou03,asghetal04,jietal05}.

In an earlier paper \citep{veras05}, we investigated whether massive planet 
scattering - while producing final dynamically settled states that admit 
stable terrestrial planets - might prevent terrestrial planets from surviving 
the transition to those states. Using test particles to follow explicitly 
the evolution of material (either formed terrestrial planets, or planetary 
embryos) in the terrestrial planet zone, we concluded from a limited 
number of realizations that there were circumstances in which the scattering 
process would impede subsequent planet formation at small radii. The 
dominant process for removing test particles was physical intrusion of 
a giant planet into the terrestrial planet region - i.e. the temporary 
existence of crossing orbits.

In this paper, we quantify statistically the extent of the anti-correlation 
which massive planet scattering should introduce between the presence 
of massive and terrestrial planets in the same system. We assume, based 
on our previous results \citep{veras05}, that terrestrial planet formation 
will be strongly suppressed exterior to the minimum periastron distance 
attained by any giant planet during the dynamical evolution, and 
calculate this minimum periastron distance from an ensemble of 500 N-body 
simulations. The appropriate initial conditions for scattering experiments 
are uncertain - here we assume that the initial states are triple planet 
point mass systems in which the planets have a realistic range of 
masses \citep{tabachnik02}. Although undoubtedly oversimplified, these 
initial conditions lead to final states which match the observationally 
determined eccentricity distribution of exoplanets reasonably well.

The outline of this paper is as follows. Section \S2 describes the motivation 
behind our initial conditions, and section \S3 briefly describes the numerical 
codes used. We describe our results in \S4, and discuss some of the implications 
in \S5.

\section{Initial Conditions}
A large number of different initial configurations of massive planets are likely 
to evolve, via gravitational scattering, toward stable arrangements that match 
the observed eccentricity distribution of extrasolar planets. Massive planet formation 
theory does not provide any clear guidance as to the expected distribution of the 
number of planets that would be formed in a single system, so it is reasonable to 
adopt any number from $N=2$ \citep{rasio96} to large $N$ \citep{papaterq01}. 
That said, for simulations that ignore gas (as ours do), it is desirable for 
internal consistency to choose initial conditions that yield instability on a 
timescale comparable to, or longer than, the dispersal timescale of the 
gaseous protoplanetary disk. \cite{wolk96} estimate this timescale observationally 
to be $\lesssim 10^5 \ {\rm yr}$, although it is not well constrained. Unfortunately 
many setups do in fact yield a significant fraction of rapidly unstable initial 
conditions (for example \cite{holman93} find a logarithmic decay of the number 
of surviving test particles in an outer Solar System simulation), though it is 
possible to at least avoid extremely short (and thus grossly inconsistent) 
instability timescales. Here, we model systems of three giant planets, with 
the motivation being that they represent the simplest systems which yield instability 
on the desired timescale without considerable fine-tuning of the initial planetary 
separations.

\subsection{Initial planetary separations}
\cite{chambers96} and \cite{marzari02} plot dynamical settling times for three-planet systems 
as functions of initial separations, while \cite{ito01} show similar plots for sets of 9 and 
14 terrestrial-like planets which are perturbed by Jupiter and Saturn.  The instability 
timescales for the three-planet systems exhibit an exponential dependence on initial 
separations, but were based on constant initial planetary separations and equal mass planets, 
simplifications not assumed in this work.  Hence, although we used such diagrams as 
guidelines for establishing our initial conditions, we performed trial simulations of 
in order to determine the appropriate instability times for our runs. Based on those 
runs, we fixed our initial inner, middle, and outer planet semimajor axes at $a_1 = 5$ 
AU, $a_2 = 6$ AU, and $a_3 = 12$ AU respectively. The ratio of semimajor axes for the 
inner pair of planets, $a_2/a_1 = 1.2$, lies just within the globally chaotic limit 
established by Wisdom (1980) for two-planet systems. Typically, these semimajor axis choices 
produce instability that causes the ejection of at 
least one planet on timescales exceeding several orbits but less than several million 
years. The system dynamically settles to a state that at least qualitatively resembles 
known multiplanet exosystems --  often one featuring two well-separated eccentric giant planets.

Since we ignore physical collisions between giant planets, our simulations are freely 
scalable -- only the ratios, and not the magnitudes, of $a_1, a_2, a_3$ have physical 
significance. Nevertheless, our choices are plausible, since giant planet formation 
close (within a few AU) to a star is unlikely, while the predicted timescale for planet 
formation via core accretion \citep{pollack96} does not vary dramatically 
for $a \approx 5 - 10 \ {\rm AU}$.

\subsection{Masses}
Simulations of gravitational scattering within two-planet systems show that the 
results are dependent on the assumed mass distribution \citep{rasio96,ford03a}. Here, 
we randomly assigned masses to each giant planet 
in the range from $0.3 M_{Jup}$ to $3 M_{Jup}$, weighed by the $1/M$ power-law distribution 
determined by \cite{tabachnik02}.  This range, the lower bound of which represents a Saturn mass, 
covers one order of magnitude, and accounts for the majority of exoplanet minimum masses so far 
discovered. Our mass distribution is weighted toward subjovian masses in order to mirror the 
actual distribution of exoplanet masses. We note that we do not account for possible systematic 
differences between the mass distribution at formation, and that determined observationally 
following (within this model) gravitational scattering, which typically ejects lower mass 
planets more frequently than high mass planets. Nor do we consider correlations between 
semimajor axis and mass at the epoch of formation.

\subsection{Eccentricities and inclinations}
Having determined the initial separations and masses of the giant planets, 
we randomized each planet's angular variables for each run.
Each planet initially resides on a near-circular orbit, with an 
eccentricity of $0.01$, and a random and negligible, but nonzero, inclination, of less than 
$0.001^{\circ}$.  Such near-zero eccentricity and inclination values are similar to those used by 
\cite{ford03a} in their two-planet simulations.  We fix our initial values so as not to introduce 
additional variables which should be considered when analyzing the results.  Regardless, the 
variance in \cite{ford03a}'s initial eccentricity and inclination values do not appear to 
drastically alter their results.

\begin{figure}
\centerline{\psfig{figure=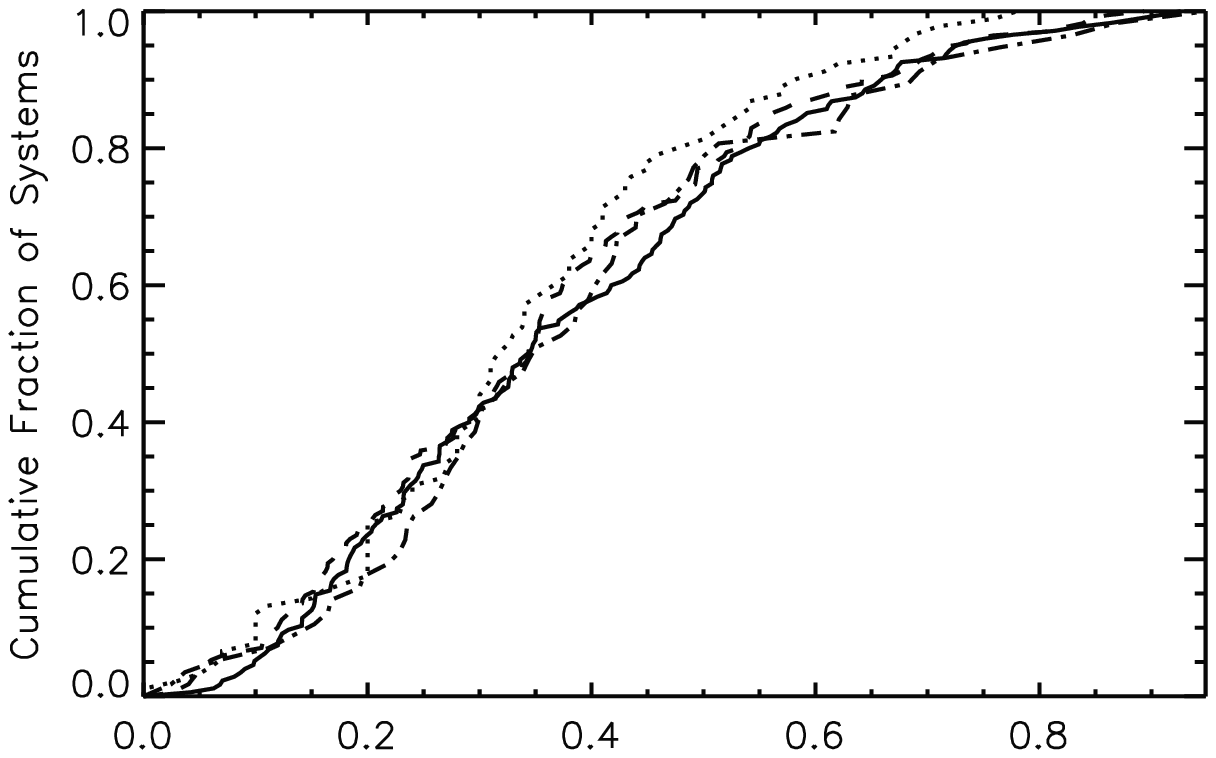,width=\columnwidth}}
\centerline{\psfig{figure=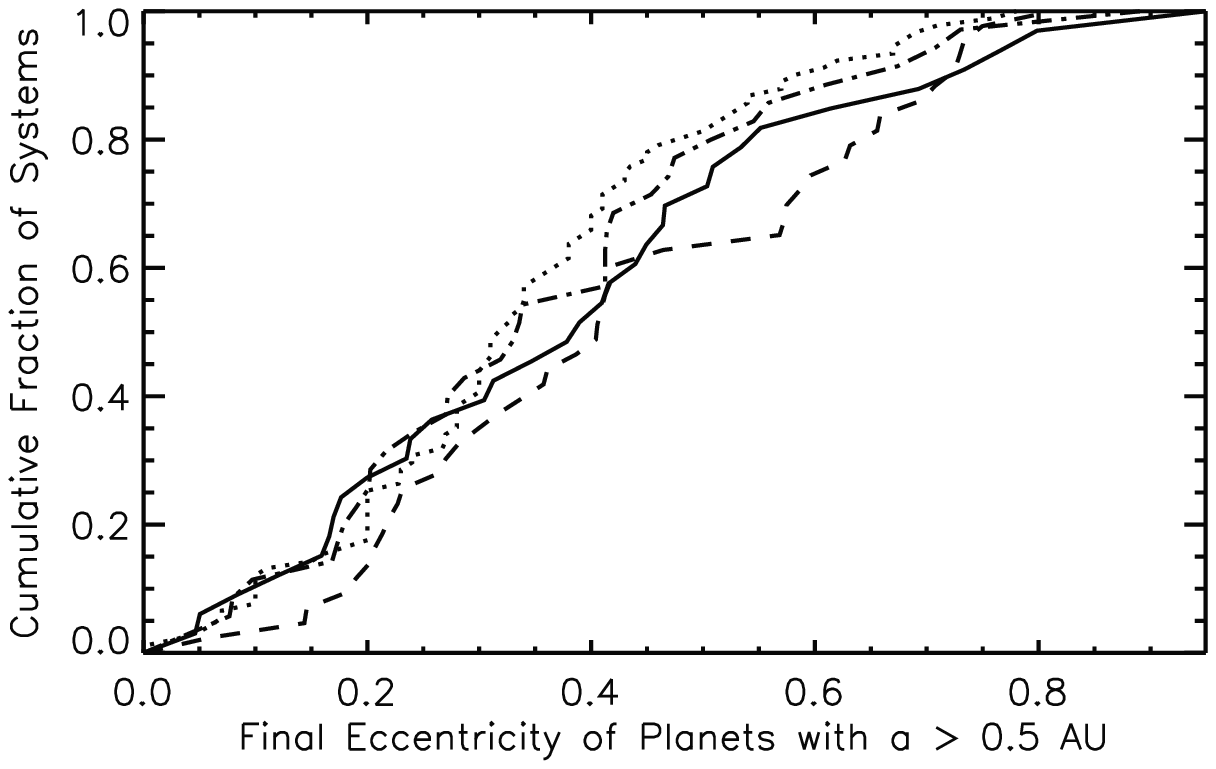,width=\columnwidth}}

\label{cume5}
\figcaption{The cumulative distribution of the observed extrasolar planet eccentricities beyond 
$a > 0.5$ AU (dotted lines) and the simulated distribution from the HNbody (solid lines) and 
Hermite (dashed lines) codes for $5$ Myr simulations (upper panel) and $10$ Myr simulations (lower
panel). The dot-dashed lines represents an additional simulated HNbody 
distribution derived from a different set of initial EGP semimajor axes of $5, 10$, and $12$ AU.}
\end{figure}

\section{Numerical Method}
We ran the scattering experiments using two independent gravitational N-body codes. Our 
primary tool was the HNbody code (K. P. Rauch \& D. P. Hamilton 2006, 
in preparation). HNBody is a collection of integrators primarily designed for systems 
with one massive central object, and has been used for both massive 
extrasolar planet simulations \citep{veras04a, veras04b, veras05} and satellite 
interactions around Neptune  \citep{zhang05}. As a check, we also ran experiments 
with a Hermite stellar dynamics code \citep{hutmakino03}, which, although not as efficient 
as HNbody for this problem, is perfectly able to evolve systems for the relatively 
modest durations of our simulations. Adjustable run-time parameters 
for HNbody include choice of integrator, accuracy parameter, and timestep, whereas only 
timestep is a required parameter for the Hermite code. Given the Bulirsch-Stoer 
integrator's ability to model accurately close encounters between planets, we 
chose this integrator for all HNbody simulations.  
The user-inputted timestep then sets the maximum timestep over which the 
integrator may iterate while still maintaining a pre-specified accuracy. We used a 
timestep of $0.025$~yr, which corresponds to a robust $1/200$ of an orbit of the 
initially innermost planet (note that during the evolution, $a$ for the innermost 
planet can decrease substantially, so a large safety margin at the start of the 
evolution is necessary). The resulting relative 
energy errors were $\sim 10^{-9} - 10^{-12}$ throughout the simulation.  For the 
Hermite code, we chose a smaller 
timestep of $1/1000$ of an orbit for the innermost planet, as the code is not 
specialized for planetary 
problems. Even with the reduction in timestep, the relative errors 
($\sim 10^{-4} - 10^{-6}$) are orders 
of magnitude larger than those from the HNbody code, but still acceptable.   
Data was outputted at every $250$-$500$ yr depending on the simulation.

\begin{deluxetable}{ccc}
\tabletypesize{\small}
\tablecaption{KS TEST NUMBERS FOR ECCENTRICITY DISTRIBUTIONS\label{KS}}
\tablecolumns{3}
\tablewidth{0pt}
\tablehead{\colhead{ } & 
   \colhead{5~Myr runs} &
   \colhead{10~Myr runs}  
}
\startdata
actual-HNbody\#1 & 13.8\%  & 36.0\%    \\
actual-Hermite & 71.5\%   & 57.6\%  \\
HNbody\#1-Hermite & 41.0\%  &  46.6\%  \\
HNbody\#2-HNbody\#1 & 88.3\%  & 96.0\%    \\
HNbody\#2-Hermite & 75.0\%   & 22.9\%  \\
HNbody\#2-actual & 75.6\%  &  41.2\% 
\enddata
\tablecomments{The KS test probability that different pairs of distributions, plotted 
in Figure 1, are actually drawn from the same parent distribution. The designation
``HNbody \#2'' represents the additional simulation run with initial EGP semimajor axes 
of $5, 10$, and $12$ AU.  All four distributions are broadly consistent.}
\end{deluxetable}

In order to maximize the number of relevant simulations run while maintaining 
acceptable standards of energy conservation, we performed a suite of preliminary tests.
As the chaotic nature of the unstable systems we analyzed produces different 
evolutionary tracks due to machine-precision roundoff error, we do not compare 
simulations run with both codes on an 
individual basis. Rather, we perform a statistical comparison. We are also interested 
in comparing the distribution of final planetary eccentricities with observations, and
for this purpose we consider only exoplanets with semimajor axes $> 0.5$ AU; closer 
planets could have been influenced by tidal effects, which were not modeled by 
our simulations.  We also only compare planets in systems in 
which dynamical instability produced crossing orbits, leaving only one or two planets 
in stable orbits.  Over half of the systems modeled with our initial conditions 
produced such instability.

\section{Results}
For our main sample, we evolved 250 systems using HNbody for 5~Myr. This duration 
is comparable (in terms of orbits) to several previous studies 
\citep{ford01,adams03,veras04a}. 
Of these runs, $243$ achieved crossing orbits, and $175$ exhibited at least one 
planet achieving a hyperbolic orbit. We also ran $250$ Hermite 5-Myr simulations, 
$244$ of which achieved crossing orbits, and $170$ of which exhibited at 
least one planet achieving hyperbolic orbits.  No planet which remained 
stable throughout our simulations evolved its semimajor 
axis to within a region where tidal effects would become important.  Therefore,
we did not need to incorporate tidal corrections into the statistical comparisons.
Although with our initial conditions significant dynamical evolution frequently 
occurs within $5$ Myr, it is by no means obvious that the ultimate dynamical state 
is attained within this period. Hence, we carried out a smaller set of $50$ 
10-Myr simulations with each integrator in order to check 
whether this alters the qualitative or quantitative results; we present results 
from both sets of simulations concurrently. The planets in all such simulations 
achieved crossing orbits. The number of systems which achieved 
hyperbolic orbits for the HNbody and Hermite simulations were $33$ and $43$ respectively. 
We only consider systems in which one or more planet has been ejected.

\begin{figure}
\centerline{\psfig{figure=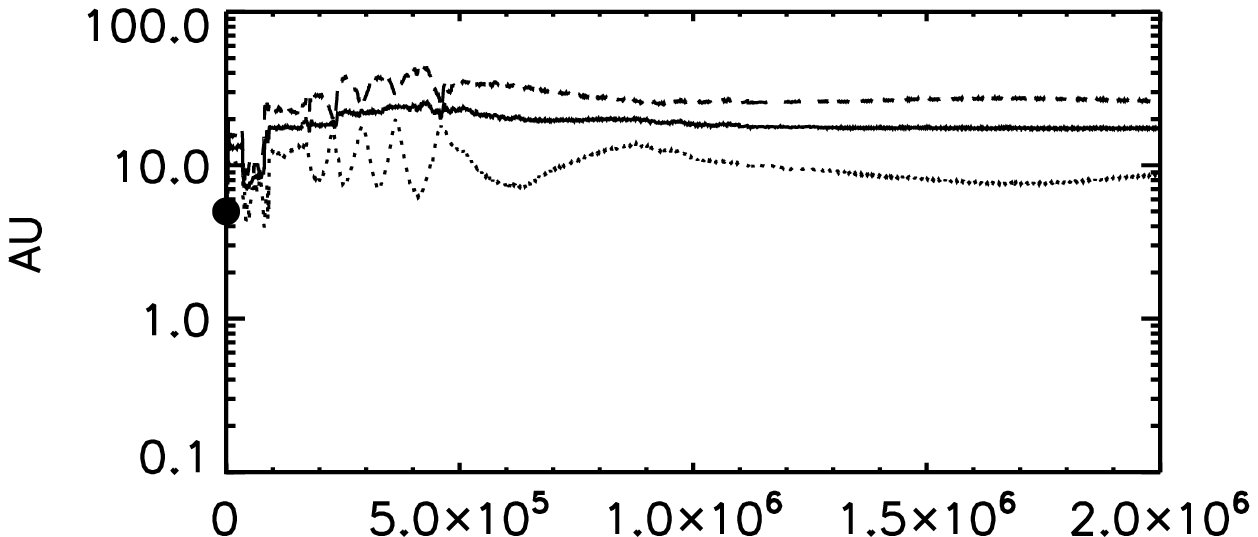,width=\columnwidth}}
\centerline{\psfig{figure=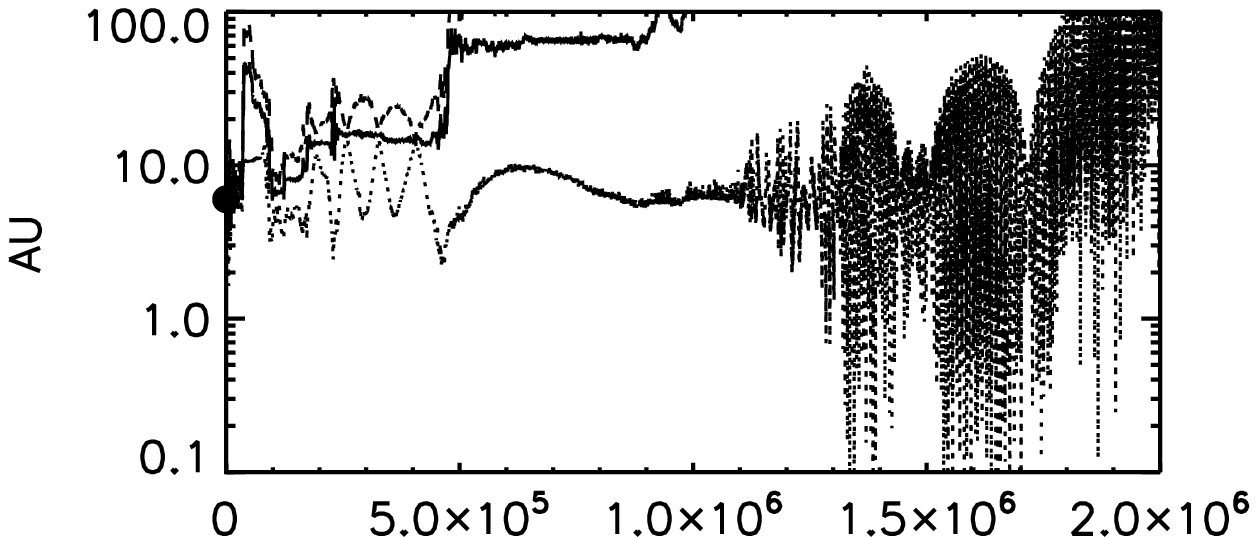,width=\columnwidth}}
\centerline{\psfig{figure=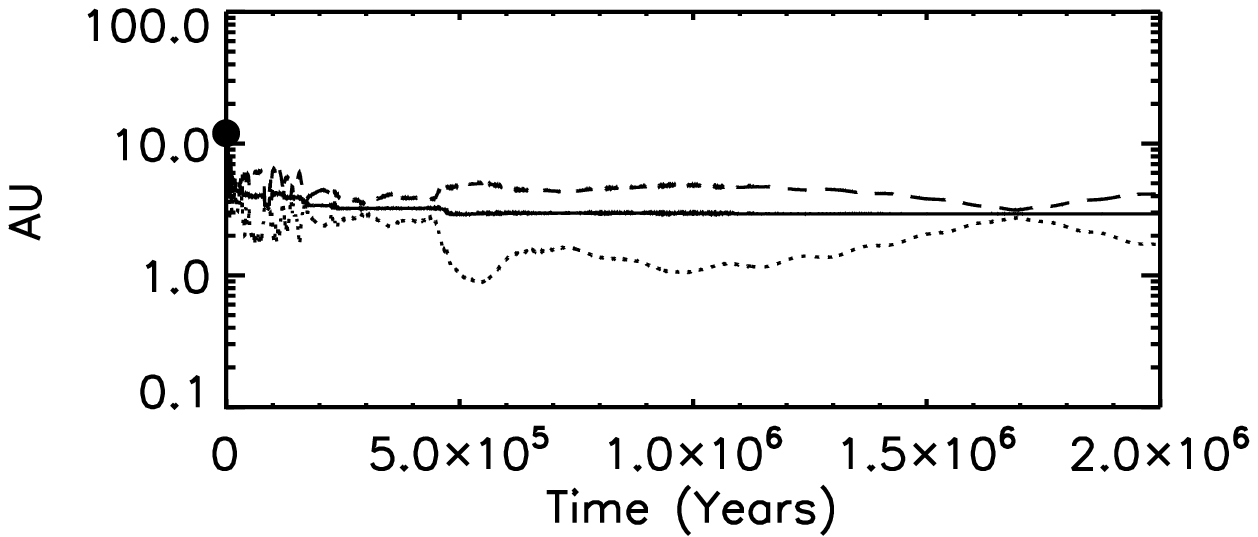,width=\columnwidth}}

\label{indplot}

\figcaption{Evolution of the semimajor axis (solid lines), periastron (dotted lines), and 
apoastron (dashed lines) of the three planets for a single run. In this case, the planet 
with $a_2 = 6$ AU at $t = 0$ (middle panel) was eliminated from the system. 
The eliminated EGP performs a nearly complete radial sweep of the system 
between $1.5$ and $2$ Myr.}
\end{figure}

Figure 1 displays the $91$ known extrasolar planet 
eccentricities as of June 1st, 2005 which satisfy $a \ge 0.5$ (dotted line), along with the
final eccentricity distributions of the surviving planets from the $5$ Myr (upper panel)
and $10$ Myr (lower panel) HNbody (solid line) and Hermite (dashed line) simulations.  
The HNbody distributions from a different set of initial EGP conditions, where $a_1 = 5$ AU,
$a_2 = 10$ AU, and $a_3 = 12$ AU, are superimposed on both panels as dash-dot lines.
Respectable agreement is obtained, both between the results 
from the two numerical codes, and between the simulations and the observed distribution 
of eccentricities of extrasolar planets. 
Quantitatively, a KS test (Table 1) verifies 
that the simulated and observed eccentricities are consistent with their being 
drawn from the same parent distribution. This agreement may be somewhat 
fortuitous, especially at the high $e$ end of the distribution, where the 
observed population may have been underestimated due to selection effects 
\citep{cumming04}. Less than $2$\% of the planets
simulated in this work attain eccentricities higher than $0.8$, the upper eccentricity
limit predicted by \cite{ford03a} for the two-planet scattering model in which both
planets initially reside on coplanar, circular orbits.  The occasional higher 
eccentricity values found in this study may be attributed to the four-body nature of the
system and the small ($= 0.01$) nonzero initial EGP eccentricity values.

\begin{figure}
\plotone{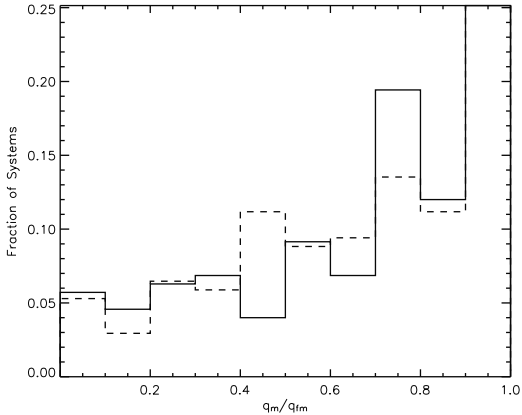} 
\label{indq5}
\figcaption{The distribution of the ratio $q_m/q_{fm}$ after 5~Myr, where $q_m$ is the minimum 
periastron distance achieved by any planet throughout the simulation, and $q_{fm}$ is the minimum 
periastron among the surviving planets at the final state of the simulation.  
HNbody simulations are represented by solid lines and Hermite simulations are represented by dashed lines.}
\end{figure}

\begin{figure}
\plotone{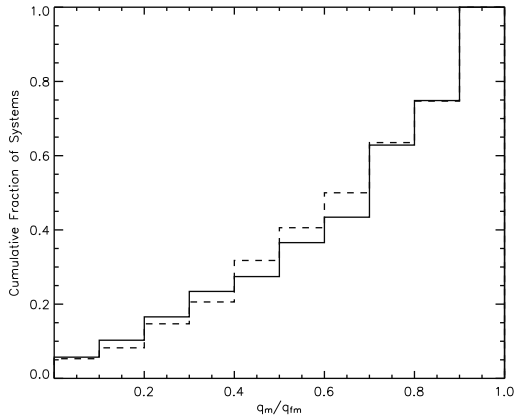} 
\label{cumq5}
\figcaption{The cumulative distribution of the ratio $q_m/q_{fm}$ after 5~Myr, where $q_m$ is 
the minimum periastron distance achieved by any planet throughout the simulation, and $q_{fm}$ is 
the minimum periastron among the surviving planets at the final state of the simulation. 
HNbody simulations are represented by solid lines and Hermite simulations are represented by dashed lines.}
\end{figure}

The agreement demonstrated in Figure 1 suggests that our simulations yield 
outcomes that are consistent with the dynamical properties of actual systems. Of 
course this agreement is in no way unique, and other scattering scenarios are known 
that produce similar results. For example, we obtain rather similar results to 
those of \cite{ford03a}, despite substantially different initial conditions. A 
range of masses among the scattering planets does appear to be important.

Within our model, the effective sphere of influence of the giant planets on 
subsequent terrestrial planet formation is dictated by the radial excursions of 
the giant planets during their dynamical evolution. Since dynamical instability 
often causes crossing orbits, the most restrictive conditions (i.e. the 
greatest inward excursions) can be set by any of the initial three planets. 
One example is depicted in Figure 2, which shows the 
evolution of the semimajor axis, pericenter and apocenter of the three 
planets for the first 2 Myr of a 5~Myr run. In this case, the final state of the system 
leaves two planets in moderately eccentric, well-separated periodic orbits. 
However, in the evolution up to this state, the initially innermost and outermost planets 
switched positions, while the third planet swept radially through the entire system within 
2~Myr, and harbored a pericenter which remained close to the parent star for nearly $0.5$ Myr.
This behavior is effectively hidden by the benign final state of the system.

\begin{figure}
\plotone{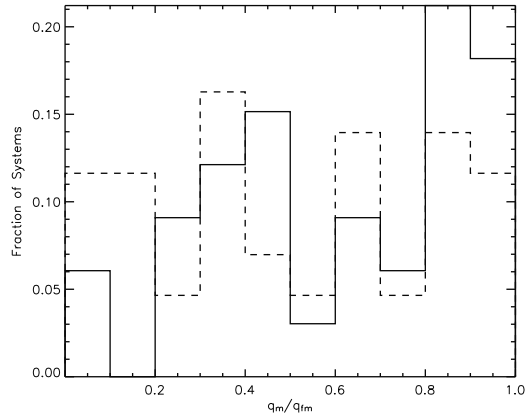} 
\label{indq10}
\figcaption{The distribution of the ratio $q_m/q_{fm}$ after 10~Myr, where $q_m$ is the minimum 
periastron distance achieved by any planet throughout the simulation, and $q_{fm}$ is the minimum 
periastron among the surviving planets at the final state of the simulation.  
HNbody simulations are represented by solid lines and Hermite simulations are represented by dashed lines.}
\end{figure}

In order to quantify how common such evolutionary histories are, we consider the ensemble of 
simulations discussed above. We denote by $q_m$ the minimum periastron distance 
achieved by any planet throughout the simulation, and by $q_{fm}$ the minimum periastron 
among the surviving planets at the termination of the simulation. Figure 3 shows the 
distribution of the ratio of these distances, $q_m/ q_{fm}$, for the $5$ Myr 
HNbody (solid line) and 
Hermite (dashed line) simulations respectively. Figure~4 is the corresponding set of cumulative 
histograms. We find, as previously \citep{veras05}, that the periastron distance for 
the innermost planet in the final, relaxed state, is often a poor proxy for the 
extent of radial sweeping that occurs during the earlier dynamical evolution. 
Although $q_m/ q_{fm}$ does increase toward  $q_m/ q_{fm} = 1$, the distribution 
is broad and extends down to quite small values $q_m/ q_{fm} \lesssim 0.2$. From 
the cumulative histograms, we find that $q_m/ q_{fm} \lesssim 0.2$ in approximately 
10\% of realizations, while $q_m/ q_{fm} \lesssim 0.5$ in approximately 30\% of 
cases. Within a scattering scenario, we would therefore expect to observe 
a significant deficit of lower mass planets in orbits with semimajor axes 
less than half the periastron distance of the innermost massive planet. Many 
orbits in this exclusion zone would be stable today, but would have been 
unfavorable for planet formation due to the early massive planet dynamics.
No significant difference between results computed with the two codes is seen. 
Broadly similar results are also obtained from the simulations run on for 10~Myr -- plotted 
in Figure 5 -- albeit with poorer statistics.

\begin{figure}
\centerline{\psfig{figure=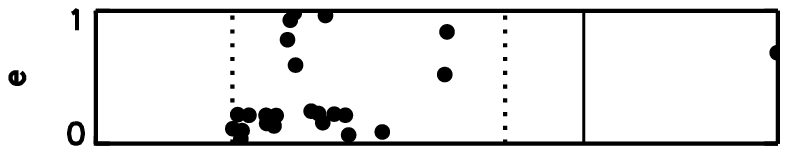,width=\columnwidth,height=0.65truein}}
\centerline{\psfig{figure=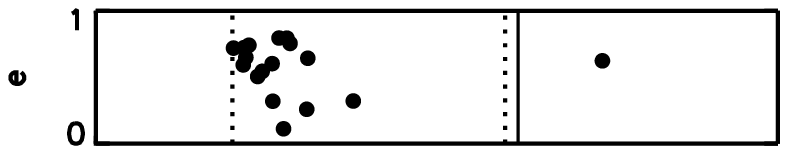,width=\columnwidth,height=0.65truein}}
\centerline{\psfig{figure=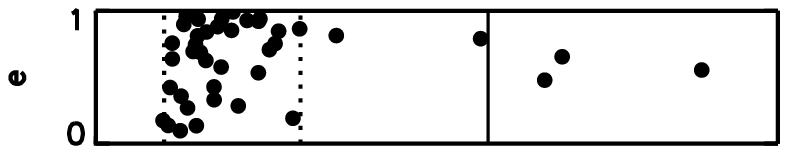,width=\columnwidth,height=0.65truein}}
\centerline{\psfig{figure=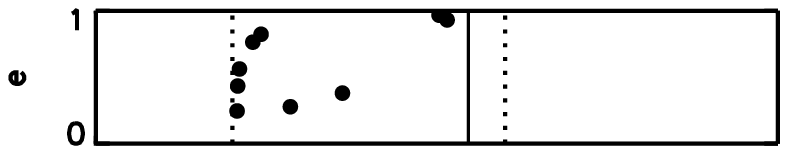,width=\columnwidth,height=0.65truein}}
\centerline{\psfig{figure=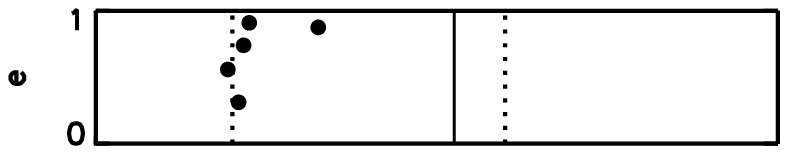,width=\columnwidth,height=0.65truein}}
\centerline{\psfig{figure=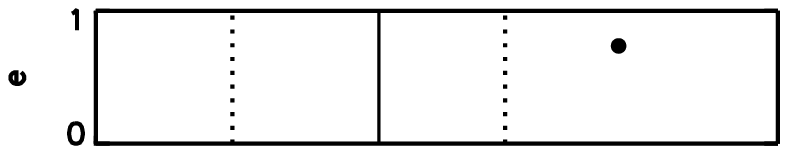,width=\columnwidth,height=0.65truein}}
\centerline{\psfig{figure=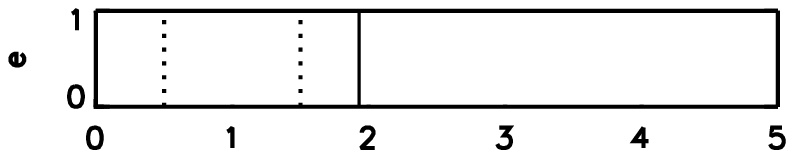,width=\columnwidth,height=0.65truein}}
\label{indplot}

\bs

\bs

\bs

\figcaption{Final states, after $5$ Myr, of $7$ simulations each run with 3 EGPs 
and 50 terrestrial planets, where dots denote surviving terrestrial planets, 
dotted lines enclose the initial semimajor axes of those planets, and solid 
lines denote $q_m$, the minimum periastron distance achieved by any EGP 
throughout the simulation.  Two different initial terrestrial planet annuli 
are sampled.}
\end{figure}

No simulations so far described include terrestrial planets.  As a further 
demonstration of our results and as a check on the approximate criterion for
strong interactions adopted earlier, we performed
7 additional simulations including 50 terrestrial planets over 5 Myr with HNbody.  
The planets were treated as ``Light Weight Particles'' (LWPs), which are test 
particles that do not perturb EGPs. The LWPs were initially
placed uniformly in annuli ranging from $0.5-1.5$ AU ($2$ simulations) or 
$1.0-3.0$ AU ($5$ simulations) on circular, nearly coplanar orbits.  Figure 6 provides
a snapshot of the final states of all $7$ simulations, with dots representing surviving
LWPs, solid lines representing $q_m$, and dotted lines enclosing the initial LWP 
annulus.  The plots are arranged according to decreasing $q_m$ in order to illustrate
just how sensitively terrestrial planet survival depends upon EGP excursion into
the inner system.  These runs show that although occasionally low mass bodies may 
survive with semi-major axes outside the minimum periastron separation, typically 
there's significant clearing even within the minimum periastron distance.
Hence, the approximate criteria we've adopted is on the conservative side.
The uppermost panel indicates that the initially low eccentricities of LWPs
will remain near-circular if the EGPs fail to intrude upon a region within twice the
radial distance of the terrestrial material; the seemingly anomalous vertical band
of LWPs at $\approx 1.3$ AU may be attributed to strong resonant interactions with the
approaching EGPs.

The random planet mass distribution used in the simulations raises an additional 
question -- which combinations of planet masses promote the greatest extent 
of radial sweeping? Under some circumstances (for example when two planets 
remain on stable bound orbits at the end of the evolution), clues as to the 
initial masses of the scattering planets may survive the scattering process, 
so any strong correlation between the masses and the outcome is potentially 
observable. Figure~7 shows how the mean $q_m$ scales with the range of planet 
masses in the randomly chosen initial conditions, quantified via the standard 
deviation of the log of the planet masses. In the figure, solid lines
denote the $5$ Myr simulations, dashed lines indicate $10$ Myr simulations, and triangles
and squares refer, respectively, to the HNbody and Hermite code simulations. 
We find a roughly linear trend -- as the scatter in the mass increases, so does the 
mean value of $q_m$. Systems in which the planets are of roughly similar masses 
are therefore most detrimental to the survival of interior terrestrial planet material. 
Given our assumed $1/M$ distribution of masses, we expect EGP standard mass deviations 
of over $1 M_{Jup}$ to be relatively rare.

\begin{figure}
\plotone{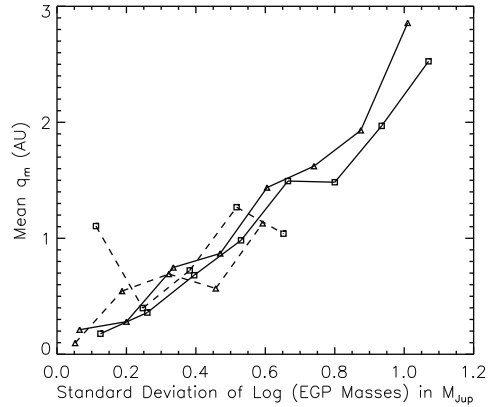} 
\label{mass5}
\figcaption{Variation of the mean $q_m$ values with the variance of the initial planet 
masses, for $5$ Myr (solid line) and $10$ Myr (dashed line) simulations using the 
HNbody (triangles) 
and Hermite (squares) codes. Plotted points refer to binned values of the logged mass
standard deviation. Masses were randomly sampled from a $1/M$ distribution.}
\end{figure}

Within our assumed scattering scenario for the origin of extrasolar planet 
eccentricities, the extent of the penetration of giant planets into the 
terrestrial planet zone is likely to be the dominant factor determining 
whether the evolution hinders subsequent terrestrial planet formation. 
However, the amount of time spent with giant planets at small periastron 
distances may also be significant. We find a wide gap between the timescale 
on which crossing orbits first develop, and the timescale on which a 
planet is first forced onto a hyperbolic orbit. Crossing orbit times for our 
simulations are on the order of $10^3$ yr -- short enough that the effects 
of a residual decaying gas disk would obviously be important and modify 
the result. By contrast, the timescale for ejections is of the order of 
a Myr. Figure~8 is a histogram of the initial time on which a planet achieves 
a hyperbolic orbit in the $5$ Myr simulations for the HNbody (solid line) and 
Hermite (dashed line) simulations. Because the timescales on which planetary orbits 
merely cross versus become hyperbolic vary by several orders 
of magnitude, Saturn or Jovian-mass planets may be dynamically excited for hundreds of 
thousands of years before being ejected from the system, suffering a collision, or 
dynamically settling into a quasi-stable configuration such as that in Figure~2. 
Intrusions into the terrestrial planet zone are therefore typically repeated and / or 
extended, rather than being one-time events.

\begin{figure}
\plotone{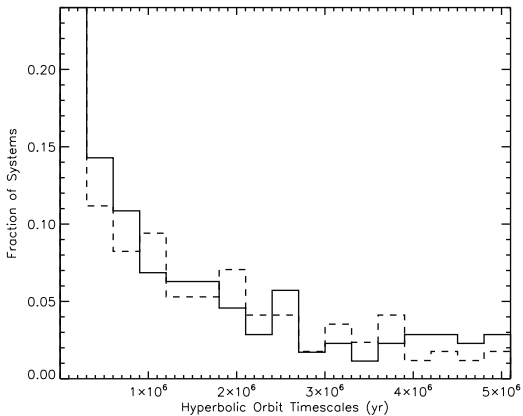} 
\label{hyper5}
\figcaption{Histogram of times at which a planet first traveled on a hyperbolic orbit. 
The planets were evolved for $5$ Myr. HNbody simulations are represented by solid lines 
and Hermite simulations are represented by dashed lines.}
\end{figure}

\section{Discussion}
Forthcoming missions that promise to detect terrestrial planets within the 
habitable zone -- such as {\em Kepler} -- will also open the possibility of studying 
statistically the influence of massive planets on terrestrial planet formation 
(at least in the limited cases in which radial velocity surveys prove sensitive 
enough to constrain the giant planet populations around typically rather 
faint stars). That the existence of Jupiter and Saturn affects the formation 
of Earth and the rest of the inner Solar System, particularly in the vicinity 
of the asteroid belt, is well-established \citep{wetherill92,lecar97,chambers02,
leviagno03,mardling04}. The distribution of extrasolar planetary masses is 
securely predicted to show a similar bimodality \citep[e.g.]{ida04}, due to the 
rapid timescale on which runaway accretion of giant planet envelopes occurs 
\citep{pollack96}, so a coupling between the observed properties of the 
giant and terrestrial planets in the same system is expected. Since giant 
extrasolar planets have substantially different orbital properties (and, 
perhaps, evolutionary histories) than those planets in the Solar System, we have 
argued here that the coupling between giant and terrestrial planet formation 
might be atypically weak in the Solar System, and more dramatic in extrasolar 
planetary systems.

The orbital properties of giant extrasolar planets immediately suggest two 
new mechanisms via which the evolution of giant planets might impede 
terrestrial planet formation. First, the orbital migration that is 
believed to be necessary to explain the origin of Hot Jupiters \citep{lin97b},
and possibly the Late Heavy Bombardment \citep{gomes05}, 
means that, at least in some systems and possibly in the typical system 
\citep{armitage02,trilling02}, giant planets move through the entire 
terrestrial planet zone during the early evolution of the disk. This 
may reduce the probability of subsequent terrestrial planet formation 
\citep{armitage03}, though the extent of any suppression will depend 
on the speed of migration (which alters the degree of shepherding via 
resonant capture) and on the surface density of planetesimals in the 
inner regions following scattering by the migrating planet 
\citep{ward95,tanaka99,mandsigu03,edgar04,raymond05b,fogg05}. EGPs 
which accrete onto the star may elongate the dynamical relaxation time
of the planetesimal density distribution, and may occur repeatedly, further
inhibiting terrestrial planet formation.  Second, 
the large mean eccentricities of known giant extrasolar planets may 
signal that the typical planetary system is initially dynamically crowded 
and unstable to eventual gravitational scattering and planetary ejection. 
If this is the origin of eccentricity, then transient incursions by the 
giant planets into the terrestrial planet during the scattering process 
can be particularly detrimental to terrestrial planet formation, since 
the scattering is likely to occur at a relatively late epoch when the 
gas disk (which would tend to recircularize scattered planetesimals) 
has been dissipated. Studies of the dynamical stability of hypothetical 
terrestrial planets in known extrasolar planetary systems, which are 
normally based on presently observed pericenter values, may then 
hide the extent to which planets have swept out terrestrial material 
earlier in the history of that system \citep{veras05}. 

In this paper, we have attempted to quantify -- within a specific 
scattering scenario -- the extent (in radius) of the anti-correlation 
this mechanism would introduce between the presence of giant and 
terrestrial planets in the same system. We find that during 
evolutionary sequences that end in plausible-looking planetary 
systems, one massive planet sweeps inward to less than half of 
the final (`stable') periastron distance approximately 30\% of 
the time. Some effect persists (at the 10\% level) down to 
orbital radii as small as 20\% of the final periastron distance. 
If these simulations reflect the physical processes taking place 
in the early history of real extrasolar planetary systems, we 
predict that terrestrial planets ought to be under-abundant 
in nominally dynamically stable orbits interior to eccentric massive 
planets. 

\acknowledgments

We thank an anonymous referee for constructive comments, and Hal Levison and
Re'em Sari for valuable discussions.  This work was supported by
NASA under grants NAG5-13207, NNG04GL01G and from the Origins of Solar Systems and 
Astrophysics Theory Programs, and by the NSF under grant AST~0407040.

\pagebreak



\pagebreak



\pagebreak


\pagebreak


\pagebreak


\pagebreak
%
%
%
%

\pagebreak


\pagebreak


\pagebreak


\end{document}